\documentclass[smallextended]{svjour3}
\usepackage[dvips]{graphicx}
\usepackage{epsfig,amssymb,amsmath,color,url}
\newcommand{\beq}{\begin{equation}}
\newcommand{\eeq}{\end{equation}}
\newcommand{\beqn}{\begin{eqnarray}}

\newcommand{\eeqn}{\end{eqnarray}}
\def\bmath#1{\mbox{\boldmath$#1$}}
\DeclareMathOperator*{\argmin}{arg\,min}

\newcommand{\beqnn}{\begin{eqnarray*}}
\newcommand{\eeqnn}{\end{eqnarray*}}

\long\def\symbolfootnote[#1]#2{\begingroup%
\def\thefootnote{\fnsymbol{footnote}}\footnote[#1]{#2}\endgroup}
\usepackage{txfonts}
\usepackage{natbib}
\bibpunct{(}{)}{;}{a}{}{,}
\journalname{Exp. Astro.}

\begin{document}
   \title{Reduced Ambiguity Calibration for LOFAR}
   \author{Sarod Yatawatta
          }

   \institute{ ASTRON, Dwingeloo, NL\\ 
          \email{yatawatta@astron.nl}
             }

\date{Draft version: Final version published on 9 April 2012}
\maketitle
\begin{abstract}

Interferometric calibration always yields non unique solutions. It is therefore essential to remove these ambiguities before the solutions could be used in any further modeling of the sky, the instrument or propagation effects such as the ionosphere.
We present a method for LOFAR calibration which does not yield a unitary ambiguity, especially under ionospheric distortions.  We also present exact ambiguities we get in our solutions, in closed form. Casting this as an optimization problem, we also present conditions for this approach to work. 
The proposed method enables us to use the solutions obtained via calibration for further modeling of instrumental and propagation effects. We provide extensive simulation results on the performance of our method. Moreover, we also give cases where due to degeneracy, this method fails to perform as expected and in such cases, we suggest exploiting diversity in time, space and frequency.
\end{abstract}
%
\section{Introduction}
Self calibration is essential for radio interferometers such as LOFAR\footnote{The Low Frequency Array: http://www.lofar.org} to obtain high quality results under the presence of corruptions. In this paper, we focus most of our attention on propagation effects caused by the ionosphere. The ionosphere is an active medium which affects electromagnetic radiation passing through it. Naturally, radio interferometric observations on the earth are distorted by its effects due to this reason. Most significant distortions were first observed in very long baseline interferometry (VLBI) due to the fact that extremely long baselines will see the sky through unrelated parts of the ionosphere. We refer the reader to \cite{Cotton95}, (and references therein) for an overview of earliest attempts of VLBI calibration under the influence of the ionosphere. Such attempts have become simpler since the introduction of the matrix measurement equation \citep{HBS}, which encapsulates the full polarized form of radio interferometric data in a concise way.

Ionospheric distortions can also become significant at very low frequencies, even with baselines of moderate length. Therefore, instruments such as LOFAR, MWA, PAPER, GMRT, and SKA which can observe at frequencies close to ionospheric cutoff will certainly have to cope with such disturbances. Significant work has already being done in calibration and correction for ionospheric distortions of interferometric observations at very low frequency and the reader is referred to e.g., \citep{Int09,Tol07}. This paper is not about tackling such distortions, but rather about extraction of ionospheric (as well as instrumental) information with minimal ambiguities. As shown by \cite{H4}, self calibration always introduces certain ambiguities of the solutions.  At this point, we shall clarify the difference between a degeneracy and an ambiguity. A degeneracy arises when we do not have enough observations to extract complete information from the data. For example, consider locating a point in three dimensional space, just by distance measurements from a set of known reference points. If we have only two distance measurements, the location of the point will have a degeneracy. That is, it could lie anywhere on the intersection of the two spheres (with radii equal to the measured distances) from the reference points. However, by having three or more properly placed reference points, we could eliminate this degeneracy. In contrast, an ambiguity arises due to our description of a problem, or due to the choice of a coordinate system to describe our parameters. Take the case of finding the square root of a given number: We know that we could always have more than one (complex) solution to this problem. Note that this cannot be eliminated by taking more measurements.  

In the case of radio interferometry, we have more than enough baselines compared to the number of stations (or unknowns) to eliminate any degeneracy. However, the ambiguities remain, regardless of the number of baselines or data points we have. This is not a major concern in most cases with an unpolarized sky as the ambiguity will not affect the end result. On the other hand, to construct  models for instrumental and propagation effects we need to use solutions obtained by calibration. In such situations, it is essential that we use ambiguity free solutions. Traditional radio telescopes such as the Giant Meterwave Radio Telescope (GMRT) and the Very Long Baseline Array (VLBA) eliminate the ambiguities by periodically using a noise injection source but this if far from practical for a telescope like LOFAR.

Calibration of a phased array interferometer such as LOFAR has its own challenges as well as advantages in comparison with traditional dish based (movable) interferometers. Due to not having any movable components, approaching electromagnetic radiation from a celestial source will have a time varying relationship with the LOFAR dipoles  because of the rotation of the sky. On the other hand, due to the same reason (no movable parts), the LOFAR beamshape is simpler to simulate using numerical electromagnetic software. In contrast, a steerable dish based telescope can track a source across the sky but the beamshape is more complex to simulate due to shadowing, spillover etc.  Numerical simulation of LOFAR element beamshape has given us extensive knowledge of its behavior. Therefore, in this paper, we shall use the a priori knowledge of the LOFAR beamshape to our fullest advantage to restrict ambiguities in calibration.

This paper is organized as follows: In section \ref{sec:cal}, we give a brief overview of LOFAR calibration and the arising ambiguities. Next, in section \ref{Amb}, we derive expressions for the ambiguities in closed form. In section  \ref{Opt}, we discuss the necessary conditions to obtain reduced ambiguity solutions and cases where these conditions will not hold. Finally, we present results based on simulations in section \ref{Res} before concluding.

Notation: Matrices are denoted in bold uppercase letters (e.g. ${\bf A}$) and vectors in bold lowercase letters (e.g. ${\bf v}$). The sets of Complex, Real and Integer numbers are denoted as ${\mathbb C}$,${\mathbb R}$ and ${\mathbb Z}$ respectively. An estimate (or a possible solution) for a given quantity (e.g. ${\bmath \theta}$) is given by $\widehat{\bmath \theta}$. The column vector with all ones is denoted by ${\bf 1}$ and all zeros by ${\bf 0}$. The submatrix created by selecting a subset of rows (or columns) from matrix ${\bf A}$ is represented as ${\bf A}_{[a:b,c:d]}$, where we have selected rows $a$ to $b$ and columns $c$ to $d$. The pseudoinverse of ${\bf A}$ is denoted by ${\bf A}^\dagger$ while the norm by $\|{\bf A}\|$.

\section{Calibration\label{sec:cal}}
Consider the correlations observed on baseline $p$-$q$:  We can express the observed visibility matrix (ignoring additive noise) for baseline $p$-$q$ in the standard \citep{HBS} Jones formalism as
\beq \label{ME}
{\bf V}_{pq}={\bf G}_p {\bf E}_p {\bf F}_p {\bf C} {\bf F}_q^H {\bf E}_q^H {\bf G}_q^H
\eeq
where we have $N$ stations and $(p,q)$ are drawn from the set
\beq
\mathcal{S}=\{(1,2),(1,3),\ldots,(1,N),\ldots,(N-1,N)\}.
\eeq
The number of baselines (cardinality of $\mathcal{S}$) is $N_b\le N(N-1)/2$.
The matrices ${\bf G}$, ${\bf E}$ and ${\bf F}$ in (\ref{ME}) correspond to the instrumental gain, element beamshape and the ionospheric Faraday rotation, (qualified by subscript $p$ or $q$ to denote the station) respectively.

We first make the following assumptions, mainly to simplify our analysis:
\begin{itemize}
\item We observe a dominant, unpolarized, point source, at the phase center (therefore no Fourier phase component) and the station beams are pointing at that source (therefore station beam gain is unity and ${\bf E}$ represents the dipole element beam).
\item We assume electronic leakage is small (therefore {\bf G} is diagonal). Moreover, if there is any leakage it is assumed to be part of the dipole {\bf E} Jones matrix.
\item We assume we have a priori knowledge of the ${\bf E}$ Jones matrix. Moreover, because all stations have parallel dipoles (as in LOFAR), the {\bf E} Jones matrices are assumed to be identical for all stations.
\item Ionospheric delay phase is absorbed into the clock delays in {\bf G} Jones.
\end{itemize}
Note that the assumptions made here are specific for a typical LOFAR observation and therefore the results based on these assumptions are mainly applicable to a telescope that is very similar to LOFAR.

The unknowns for station $p$ are given in ${\bf G}_p$ and ${\bf F}_p$:
\beq \label{GF}
{\bf G}_p= \left[ \begin{array}{cc}
{g}_{xp}& 0\\
0& {g}_{yp}\\
\end{array} \right],\ 
{\bf F}_p= \left[ \begin{array}{cc}
\cos(\alpha_p)& \sin(\alpha_p)\\
-\sin(\alpha_p)& \cos(\alpha_p)\\
\end{array} \right],\
\eeq
where $g_{xp},g_{yp}\in {\mathbb C}$ and $\alpha_p \in {\mathbb R}$ are the unknown station gain and Faraday rotation parameters, respectively.
The element beam shape determines the ${\bf E}$ ($={\bf E}_p={\bf E}_q$) Jones matrices, about which we have fairly accurate knowledge from numerical simulations.
The source coherency (with intensity $I$) is given as
\beq \label{C}
{\bf C}= \left[ \begin{array}{cc}
I& 0\\
0& I\\
\end{array} \right].
\eeq

Consider the problem of calibration: We can represent (\ref{ME}) as
\beq \label{JME}
{\bf V}_{pq}= {\bf J}_p {\bf C} {\bf J}_q^H
\eeq
where ${\bf J}_p$ and ${\bf J}_q$ are the Jones matrices that we need to estimate. However, we can get solutions for (\ref{JME}) such as $\widehat{\bf J}_p={\bf J}_p {\bf U}$, where ${\bf U}$ is an unknown unitary matrix. This is perfectly fine if we only need to correct the data using these solutions because the unitary ambiguity cancels out in the final product. However, if we need to study the parameters in (\ref{GF}), it is hard to disentangle the unitary ambiguity from the solutions.

Another way of looking at this problem is by counting the number of degrees of freedom that is actually present in the problem. If we consider the unknowns $g_{xp}$,$g_{yp}$ (4 real numbers), $\alpha_p$ (1 real number), we get $5$ degrees of freedom per station. However, if we attempt calibration as in (\ref{JME}), we have $8$ degrees of freedom, which is clearly an overkill.

Therefore, instead of solving for a system as in (\ref{JME}), we use the original form (\ref{ME}) with unknown parameters as given in (\ref{GF}). Still, this does not guarantee us an ambiguity free solution as we see later. Hence, we study the possible ambiguities of (\ref{ME}) in section \ref{Amb} and give the necessary conditions to get a solution in section \ref{Opt}.

\section{Ambiguity Analysis \label{Amb}}
Consider the unknowns $g_{xp},g_{yp}$ in (\ref{GF}). We can rewrite this as $g_{xp}=|g_{xp}|\exp(j\phi_{xp})$, $g_{yp}=|g_{yp}|\exp(j\phi_{yp})$ where $\phi_{xp},\phi_{yp}$ are the unknown (true) phases. In order to study the ambiguities we could get for the unknowns in (\ref{GF}), we could rewrite the possible solutions as
\beq \label{aG}
\widehat{\bf G}_p= \left[ \begin{array}{cc}
(-1)^{k_{xp}}|{g}_{xp}| \times & 0\\
 \exp\left(j (\widehat{\phi}_{xp} + k_{xp} \pi + k_{\alpha p}\pi)\right)& \\
& (-1)^{k_{yp}}|{g}_{yp}|\times \\
0& \exp\left(j (\widehat{\phi}_{yp} + k_{yp} \pi + k_{\alpha p}\pi)\right)\\
\end{array} \right], 
\eeq
and
\beq \label{aF}
\widehat{\bf F}_p= \left[ \begin{array}{cc}
\cos({\widehat \alpha}_p + k_{\alpha p}\pi)& \sin( {\widehat \alpha}_p + k_{\alpha p}\pi)\\
-\sin( {\widehat \alpha}_p + k_{\alpha p}\pi)& \cos( {\widehat \alpha}_p + k_{\alpha p}\pi)\\
\end{array} \right],
\eeq
where $k_{xp},k_{yp},k_{\alpha p}\in {\mathbb Z}$. Note that if we substitute (\ref{aG}) and (\ref{aF}) instead of (\ref{GF}) into (\ref{ME}), we should still get the same value for ${\bf V}_{pq}$. The necessary conditions for us to rewrite (\ref{GF}) as (\ref{aG}) and (\ref{aF}) will be discussed in section \ref{Opt}.

Consider the product
\beq \label{EE}
{\bf E}_{pq}\buildrel\triangle\over={\bf E}_p {\bf F}_p {\bf C} {\bf F}_q^H {\bf E}_q^H = \left[ \begin{array}{cc}
E_{11,pq}& E_{12,pq}\\
E_{21,pq}& E_{22,pq}\\
\end{array}  \right]
\eeq
such that ${\bf V}_{pq}={\bf G}_p{\bf E}_{pq}{\bf G}_q^H$. Assume ${\bf E}_{pq}$ to have non zero off diagonal elements i.e., $E_{12,pq}\ne 0$, $E_{21,pq}\ne 0$. The necessary conditions for this is related to the same conditions for (\ref{aG}) and (\ref{aF}) to be valid solutions to (\ref{ME}) and will be given in section \ref{Opt}.

Then, in order to get the same ${\bf V}_{pq}$ in (\ref{ME}), by substitution of ${\bf G}_p,{\bf G}_q$ given in (\ref{GF}) as well as ${\widehat{\bf G}_p},{\widehat{\bf G}_q}$ given in (\ref{aG}) to (\ref{ME}), we get the following conditions:
\beqn \label{ke}
k_{xp}+k_{yp}=0,2,\ldots\\\nonumber
k_{xp}+k_{yq}=0,2,\ldots\\\nonumber
k_{yp}+k_{xq}=0,2,\ldots\\\nonumber
k_{yp}+k_{yq}=0,2,\ldots
\eeqn
and
\beqn \label{phie}
\widehat{\phi}_{xp}-\widehat{\phi}_{xq}+k_{xp}\pi-k_{xq}\pi+k_{\alpha p}\pi-k_{\alpha q}\pi = \phi_{xp}-\phi_{xq}\\\nonumber
\widehat{\phi}_{xp}-\widehat{\phi}_{yq}+k_{xp}\pi-k_{yq}\pi+k_{\alpha p}\pi-k_{\alpha q}\pi = \phi_{xp}-\phi_{yq}\\\nonumber
\widehat{\phi}_{yp}-\widehat{\phi}_{xq}+k_{yp}\pi-k_{xq}\pi+k_{\alpha p}\pi-k_{\alpha q}\pi = \phi_{yp}-\phi_{xq}\\\nonumber
\widehat{\phi}_{yp}-\widehat{\phi}_{yq}+k_{yp}\pi-k_{yq}\pi+k_{\alpha p}\pi-k_{\alpha q}\pi = \phi_{yp}-\phi_{yq}.
\eeqn
We can write similar condition for all other baselines.

Accumulating all possible constraints such as (\ref{ke}) for all baselines, we have
\beq \label{Ai}
\mathop{\underbrace{\bf{A}_k}_{4 N_b\times 2 N}} \mathop{\underbrace{\bf k}}_{2 N\times 1}=\mathop{\underbrace{\bf n}}_{4 N_b\times 1}
\eeq
where ${\bf k}\buildrel\triangle\over =[k_{x1},k_{y1},k_{x2},k_{y2}\ldots,k_{y N}]^T$. The vector ${\bf n}$ has only even integers as entities. This is a linear Diophantine equation (because ${\bf k} \in {\mathbb Z}^{2N}$). The matrix ${\bf A}_k$ has ones at two locations on each row. We can show this has rank of $2N$ (full column rank) using similar techniques as in \citep{Chen01,vehi03}. Therefore, we get a unique solution to ${\bf k}$ (ideally equal to ${\bf 0}$).

Next, we can also accumulate equations in (\ref{phie}) for all possible baselines to arrive at
\beq \label{Ap}
\mathop{\underbrace{\bf{A}_p}_{4 N_b\times 2 N}}\left( \mathop{\underbrace{{\widehat{\bmath \phi}}}_{2 N\times 1}} + \mathop{\underbrace{{{\bf k}}}_{2 N\times 1}}\pi + \mathop{\underbrace{{{\bf k}_{\alpha}}}_{2 N\times 1}}\pi \right) =\mathop{\underbrace{\bf{A}_p}_{4 N_b\times 2 N}}\mathop{\underbrace{{{\bmath \phi}}}_{2 N\times 1}}
\eeq
where $\widehat{\bmath \phi}\buildrel\triangle\over=[\widehat{\phi}_{x1},\widehat{\phi}_{y1},\widehat{\phi}_{x2},\widehat{\phi}_{y2},\ldots,\widehat{\phi}_{Ny}]^T$,  ${\bmath \phi}\buildrel\triangle\over=[{\phi_{x1}},{\phi_{y1}},{\phi_{x2}},{\phi_{y2}},\ldots,{\phi_{Ny}}]^T$, and ${\bf k}_{\alpha}\buildrel\triangle\over=[k_{\alpha 1},k_{\alpha 1},k_{\alpha 2},k_{\alpha 2},\ldots,k_{\alpha N}]^T$.
The matrix ${\bf A}_p$ has $1$ and $-1$ at two locations on each row.
Similar as in \citep{Chen01,vehi03}, we can show that ${\bf A}_p$ has rank $2N-1$ (null space of rank 1). Therefore, in order to find a solution to (\ref{Ap}), we have to fix one element in $\widehat{\bmath \phi}$, say by making $ \widehat{\bmath \phi}_{[1]}=\widehat{\phi}_{x1}=\phi_0$. Then, we can rewrite (\ref{Ap}) as
\beqn
\lefteqn{{\bf A}_{p[:,2:2N]}\left(\widehat{\bmath \phi}_{[2:2N]}+{\bf k}_{[2:2N]}\pi+{\bf k}_{\alpha[2:2N]}\pi\right)}&&\\\nonumber
&&={\bf A}_{p[:,2:2N]}{\bmath \phi}_{[2:2N]}-{\bf A}_{p[:,1]}\left(-{\bmath \phi}_{[1]}+\phi_0+{\bf k}_{[1]}\pi+{\bf k}_{\alpha[1]}\pi\right)
\eeqn
which gives us the solution
\beqn \label{phiest}
\lefteqn{\widehat{\bmath \phi}_{[2:2N]}={\bmath \phi}_{[2:2N]}}&&\\\nonumber
&&-{\bf A}^{\dagger}_{p[:,2:2N]}{\bf A}_{p[:,1]}\left(-{\bmath \phi}_{[1]}+\phi_0+{\bf k}_{[1]}\pi+{\bf k}_{\alpha[1]}\pi\right)\\\nonumber
&&-\left({\bf k}_{[2:2N]}\pi+{\bf k}_{\alpha[2:2N]}\pi\right).
\eeqn
To simplify (\ref{phiest}), we need to find ${\bf A}^{\dagger}_{p[:,2:2N]}{\bf A}_{p[:,1]}$. We know that the sum of each row of ${\bf A}_p$ is zero. Therefore,
\beqn \label{ones}
{\bf A}_p {\bf 1}={\bf 0}\\\nonumber
{\bf A}_{p[:,2:2N]}{\bf 1}+{\bf A}_{p[:,1]} 1={\bf 0}\\\nonumber
{\bf A}^{\dagger}_{p[:,2:2N]}{\bf A}_{p[:,1]}=-{\bf 1}.
\eeqn

Therefore, we can rewrite (\ref{phiest}) as
\beqn \label{phiest1}
\lefteqn{\widehat{\bmath \phi}_{[2:2N]}={\bmath \phi}_{[2:2N]}+{\bf 1}\left(-{\bmath \phi}_{[1]}+\phi_0+{\bf k}_{[1]}\pi+{\bf k}_{\alpha[1]}\pi\right)}&&\\\nonumber
&&-\left({\bf k}_{[2:2N]}\pi+{\bf k}_{\alpha[2:2N]}\pi\right).
\eeqn

Closer scrutiny of (\ref{phiest1}) reveals that the estimated phase, $\widehat{\bmath \phi}_{[2:2N]}$ is equal to the true phase, ${\bmath \phi}_{[2:2N]}$ with a constant unknown phase (common to all stations) and an ambiguity of $k\pi$, where $k \in {\mathbb Z}$. Therefore, the ambiguity of the solutions for the $X$ and $Y$ phases for a given station is the same (within $\pi$). The difference between the $X$ phase and $Y$ phase ambiguities is called the phase zero difference in traditional calibration jargon. Thus, we get no phase zero difference (except an ambiguity in $\pi$) for all the stations. 

Next, we shall investigate the conditions for ${\bf F}_p,{\bf F}_q$ and $\widehat{\bf F}_p,\widehat{\bf F}_q$ to satisfy (\ref{ME}) to get the same ${\bf V}_{pq}$. We arrive at equations similar to
\beq \label{alphae}
\widehat{\alpha}_p-\widehat{\alpha}_q+k_{\alpha p}\pi-k_{\alpha q}\pi = \alpha_p-\alpha_q
\eeq
for each baseline.
Combining all possible equations (\ref{alphae}) for all baselines, we get
\beq \label{Af}
\mathop{\underbrace{{\bf A}_f}_{N_b\times N}}\left( \mathop{\underbrace{\widehat{\bmath \alpha}}}_{N\times 1} + \mathop{\underbrace{{{\bf k}_{\tilde{\alpha}}}}_{N\times 1}}\pi \right) =\mathop{\underbrace{{\bf A}_f}_{N_b\times N}}\mathop{\underbrace{{\bmath \alpha}}}_{N\times 1}
\eeq
where $\widehat{\bmath \alpha}\buildrel\triangle\over=[\widehat{\alpha}_1,\widehat{\alpha}_2,\ldots,\widehat{\alpha}_N]^T$, ${\bmath \alpha}\buildrel\triangle\over=[{\alpha}_1,{\alpha}_2,\ldots,{\alpha}_N]^T$ and ${\bf k}_{\tilde{\alpha}}\buildrel\triangle\over=[k_{\alpha 1},k_{\alpha 2},\ldots,k_{\alpha N}]^T$. The matrix ${\bf A}_f$ has $1$ and $-1$ at two locations on each row. We can once again show that the rank of this matrix is $N-1$. Hence, in order to get a solution, we fix one element of $\widehat{\bmath \alpha}$, say $\widehat{\bmath \alpha}_{[1]}=\widehat{\alpha}_1=\alpha_0$. Then, we can rewrite (\ref{Af}) as
\beqn
\lefteqn{{\bf A}_{f[:,2:N]} \left(\widehat{\bmath \alpha}_{[2:N]}+{\bf k}_{\tilde{\alpha}[2:N]}\pi\right) ={\bf A}_{f[:,2:N]} {\bmath \alpha}_{[2:N]}}&&\\\nonumber 
&&- {\bf A}_{f[:,1]}\left(-{\bmath \alpha}_{[1]}+\alpha_0+{\bf k}_{\tilde{\alpha}[1]}\pi \right)
\eeqn
to yield a solution as
\beq\label{alphaest}
\widehat{\bmath \alpha}_{[2:N]}={\bmath \alpha}_{[2:N]}- {\bf A}^{\dagger}_{f[:,2:N]} {\bf A}_{f[:,1]}\left(-{\bmath \alpha}_{[1]}+\alpha_0+{\bf k}_{\tilde{\alpha}[1]} \pi \right) -{\bf k}_{\tilde{\alpha}[2:N]}\pi.
\eeq
Once again, because the sum of each row in ${\bf A}_f$ is zero, using similar analysis as (\ref{ones}), we can show that ${\bf A}^{\dagger}_{f[:,2:N]} {\bf A}_{f[:,1]}=-{\bf 1}$. Therefore, we can simplify the solution as
\beq \label{alphaest1}
\widehat{\bmath \alpha}_{[2:N]}={\bmath \alpha}_{[2:N]}+{\bf 1}\left(-{\bmath \alpha}_{[1]}+\alpha_0+{\bf k}_{\tilde{\alpha}[1]} \pi \right) -{\bf k}_{\tilde{\alpha}[2:N]}\pi.
\eeq
Therefore, our estimate $\widehat{\bmath \alpha}_{[2:N]}$ is equal to the true values, ${\bmath \alpha}_{[2:N]}$ plus an unknown constant which is common to all solutions with an ambiguity of $k\pi$, $k\in {\mathbb Z}$. Thus, we can estimate the differential Faraday rotation within a common unknown rotation and an integer ambiguity of $\pi$.

\section{Necessary Conditions \label{Opt}}
In this section, we give necessary conditions for our approach to work, especially with respect to the properties of the element beam pattern or the ${\bf E}$ Jones matrix.
\subsection{Non diagonal ${\bf E}_{pq}$ in (\ref{EE})}
First, we study the requirements for (\ref{EE}) to be non diagonal. We express the ${\bf E}$ and ${\bf F}$ Jones matrices in (\ref{EE}) as
\beqn \label{EJ}
{\bf E}_p={\bf E}_q= \left[ \begin{array}{cc}
E_{x\theta}& E_{x\phi}\\
E_{y\theta}& E_{y\phi}\\
\end{array}  \right],\\\nonumber
{\bf F}_p{\bf C}{\bf F}_q^H= I \left[ \begin{array}{cc}
\cos(\alpha_p-\alpha_q)& \sin(\alpha_p-\alpha_q)\\
-\sin(\alpha_p-\alpha_q)& \cos(\alpha_p-\alpha_q)\\
\end{array}  \right],
\eeqn
where $E_{x\theta},E_{x\phi},E_{y\theta},E_{y\phi}$ are the field components in the directions $\theta,\phi$ in cylindrical polar coordinates. The conditions required for the off diagonal terms to be zero then becomes:
\beqn \label{diagE}
&&\cos(\alpha_p-\alpha_q)(E_{x\theta}E^{\star}_{y\theta}+ E_{x\phi}E^{\star}_{y\phi})\\\nonumber
&&+\sin(\alpha_p-\alpha_q)(E_{x\theta}E^{\star}_{y\phi}-E_{x\phi}E^{\star}_{y\theta})=0\\\nonumber
&&\cos(\alpha_p-\alpha_q)(E^{\star}_{x\theta}E_{y\theta}+ E^{\star}_{x\phi}E_{y\phi})\\\nonumber
&&+\sin(\alpha_p-\alpha_q)(-E^{\star}_{x\theta}E_{y\phi}+E^{\star}_{x\phi}E_{y\theta})=0.
\eeqn
Note also that the off diagonal terms of the product ${\bf E}{\bf E}^H$ is equal to $E_{x\theta}E^{\star}_{y\theta}+ E_{x\phi}E^{\star}_{y\phi}$ (and its conjugate). If this term is zero, we get a solution to (\ref{diagE}) as $\alpha_p=\alpha_q+2k\pi$. On the other hand, if $E_{x\theta}E^{\star}_{y\phi}-E_{x\phi}E^{\star}_{y\theta}=0$, then we get another solution to (\ref{diagE}) as $\alpha_p=\alpha_q+(2k+1)\pi/2$. So, if either of these conditions are satisfied by the element beam, our method fails to give a unique solution. 

A more general way of looking at (\ref{diagE}) is as follows: We can represent (\ref{diagE}) in matrix form as
\beq\label{matdiagE}
\widetilde{\bf E}{\bf c}={\bf 0}
\eeq
where
\beqn \label{matdiagE1}
 \widetilde{\bf E}\buildrel\triangle\over=\left[ \begin{array}{cc}
(E_{x\theta}E^{\star}_{y\theta}+ E_{x\phi}E^{\star}_{y\phi})& (E_{x\theta}E^{\star}_{y\phi}-E_{x\phi}E^{\star}_{y\theta}) \\
(E_{x\theta}E^{\star}_{y\theta}+ E_{x\phi}E^{\star}_{y\phi})^{\star}& -(E_{x\theta}E^{\star}_{y\phi}-E_{x\phi}E^{\star}_{y\theta})^{\star} \\
\end{array}  \right],\\\nonumber
{\bf c}\buildrel\triangle\over=\left[ \begin{array}{c}
\cos(\alpha_p-\alpha_q)\\
\sin(\alpha_p-\alpha_q)\\
\end{array}  \right].
\eeqn
We can find a solution for ${\bf c}$ in (\ref{matdiagE}) if at least one of the eigenvalues of $\widetilde{\bf E}$ is zero (and the solution for ${\bf c}$ is the scaled eigenvector). Therefore, when $\widetilde{\bf E}$ has at least one zero eigenvalue, it is possible (under certain values of Faraday rotation) that our method fails. On the other hand, when none of the eigenvalues are zero, we are guaranteed that regardless of the Faraday rotation, our method works.

\subsection{Convex Optimization}
Normally, the solutions for the calibration are found using a non linear optimization technique. Therefore, its natural to consider the solution of (\ref{ME}) as an optimization problem. The residual error of (\ref{ME}) can be given as
\beq \label{res}
{\bf r}_{pq}=vec({\bf V}_{pq}) - {\bf f}_{pq}({\bmath \theta})
\eeq
where ${\bf f}_{pq}({\bmath \theta})=vec\left({\bf G}_p {\bf E}_p {\bf F}_p {\bf C} {\bf F}_q^H {\bf E}_q^H {\bf G}_q^H\right)$. The unknown parameter vector is given by ${\bmath \theta}$. For each station we have $5$ (real) unknowns, i.e., $|g_{xp}|,|g_{yp}|,\phi_{xp},\phi_{yp}$ and ${\alpha_p}$. However, since we can only find differential Faraday rotation (not the absolute rotation), we keep ${\alpha}$ for one station fixed. So, the parameter vector can be given as ${\bmath \theta}\buildrel\triangle\over=[|g_{x1}|,|g_{x2}|,\ldots,|g_{y1}|,|g_{y2}|,\ldots,\phi_{x1},\phi_{x2},\ldots,\phi_{y1},\phi_{y2},\ldots,\alpha_2,\alpha_3,\ldots]^T$. Hence, for $N$ stations, the size of ${\bmath \theta}$ is $5N-1$.
Finally, we can write the calibration problem as
\beq \label{theta}
\widehat{\bmath \theta}=\argmin_{\theta}\sum_{p,q} {\bf r}^H_{pq}{\bf r}_{pq}.
\eeq

The Hessian of this optimization problem will have rank $5N-2$ at the solution. This is because, we still have an ambiguity in $\phi_{xp}$ or $\phi_{yp}$. Therefore, the smallest eigenvalue of the Hessian will be zero (or close to zero under noise). For a well formulated optimization problem, we need to show that (\ref{theta}) is convex \citep{CVX} with respect to ${\bmath \theta}$. This implies that the Hessian should be positive (semi) definite. In other words, all eigenvalues (barring the smallest eigenvalue which is zero) should be positive. Therefore, by studying the second smallest eigenvalue, we can study the performance of our calibration, in particular with variation of the beam in different directions of the sky. The more positive this is, the more well formulated our problem becomes.

Under noise, the Cramer-Rao bound is inversely proportional to the second smallest eigenvalue that we discussed. Therefore, we can study the same behavior by using the Hessian instead of the Cramer-Rao bound in this case.

\subsection{Fr\'{e}chet Derivative and Condition Number}
We have assumed a priori knowledge of ${\bf E}$ in our analysis. However, there will certainly be small variations in the real beamshape from our assumed model. We need to show that small changes in ${\bf E}$ in (\ref{ME}) only leads to small changes in ${\bf V}$. This ensures that the proposed method works, even when the real ${\bf E}$ will be slightly different from our model and moreover, when each station will have a slightly different beamshape.

We use the Fr\'{e}chet derivative of (\ref{ME}) to examine the sensitivity of ${\bf V}$ to small changes in ${\bf E}$.  The Fr\'{e}chet derivative is a generalization of the {\em derivative} of scalar functions to higher dimensional functional spaces. More details about the Fr\'{e}chet derivative and sensitivity analysis of matrix functions can be found in \cite{higham}. 
The dependence of ${\bf E}$ on ${\bf V}_{pq}$ in (\ref{ME}) can be written as
\beq
{\bf V}({\bf E})={\bf V}_{pq}={\bf G}_p {\bf E} {\bf F}_p {\bf C} {\bf F}_q^H {\bf E}^H {\bf G}_q^H
\eeq
Let ${\bf \Delta E}$ denote a small change to ${\bf E}$. Then, as in \cite{higham}, we write the Fr\'{e}chet derivative as
\beq \label{Frechder}
L_{\bf V}({\bf E},{\bf \Delta E})={\bf V}({\bf E}+{\bf \Delta E})-{\bf V}({\bf E})-o(\|{\bf \Delta E}\|)
\eeq
where 
\beq
\lim_{\|{\bf \Delta E}\| \rightarrow 0} \frac{o(\|{\bf \Delta E}\|)}{\|{\bf \Delta E}\|} = 0.
\eeq
Therefore, we get the Fr\'{e}chet derivative as
\beq \label{Frech}
L_{\bf V}({\bf E},{\bf \Delta E}) = {\bf G}_p {\bf E} {\bf F}_p {\bf C} {\bf F}_q^H {\bf \Delta E}^H {\bf G}_q^H +   {\bf G}_p {\bf \Delta E} {\bf F}_p {\bf C} {\bf F}_q^H {\bf E}^H {\bf G}_q^H.
\eeq
The relative condition number of ${\bf V}({\bf E})$ is given as
\beq
cond({\bf V}({\bf E})) =\max_{\|{\bf \Delta E}\| \ne 0} \frac{\|L_{\bf V}({\bf E},{\bf \Delta E})\|}{\|{\bf \Delta E}\|} \frac{\|{\bf E}\|}{\|{\bf V}({\bf E})\|}.
\eeq
Using norm inequalities $\|{\bf AB}\|\le \|{\bf A}\|\|{\bf B}\|$ and $\|{\bf A}+{\bf B}\|\le \|{\bf A}\|+\|{\bf B}\|$ we get an upper bound for the relative condition number as
\beq \label{cond}
cond({\bf V}({\bf E})) \le \|{\bf G}_p\|(\| {\bf F}_p {\bf C} {\bf F}_q^H {\bf E}^H \| + \| {\bf E} {\bf F}_p {\bf C} {\bf F}_q^H \| ) \| {\bf G}_q \| \frac{\|{\bf E}\|}{\|{\bf V}({\bf E})\|}.
\eeq
We can use (\ref{cond}) to study the sensitivity of the data to variations in ${\bf E}$ and hence, the sensitivity of our calibration to errors in our knowledge of ${\bf E}$. Generally, it is better to have a condition number as small as possible, the smallest being equal to $1$.

\subsection{Diversity}
In this section, we have given several ways of looking at the necessary conditions for our approach to work. In practice, there will always be cases where some of these conditions are not satisfied. However, even if these conditions are not satisfied, we can exploit diversity to make our method work. Diversity can be described as the opposite of degeneracy. Indeed, degeneracy arises when we do not have enough independent observations. In contrast, when we do have observations taken independently from one another we create diversity. There are several different forms of diversity that we can exploit. Implicitly, we used the knowledge of ${\bf E}$ and moreover, the fact that this is not an identity matrix, in our proposed method. In other words, we used polarization diversity arising due to  $X$ and $Y$ dipoles having different beam patterns. Furthermore, even if the aforementioned conditions are  not satisfied for one baseline, there are other baselines where the ionosphere is different and a solution is possible. Secondly, ionosphere will almost always change with time and even for a baseline which has degeneracy, with time, things will change to yield a solution. Thirdly, the element beam changes with direction in the sky and as the source moves across the beam, it will  overcome degenerate cases. Moreover, ionospheric properties have well known frequency behavior. In situations where we fail to obtain a solution, we can extrapolate solutions obtained at different times,baselines,frequencies etc., to get a complete solution.
\section{Simulation Results\label{Res}}
\begin{figure}[htbp]
\begin{minipage}{0.99\linewidth}
\begin{minipage}{0.50\linewidth}
\centering
 \centerline{\epsfig{figure=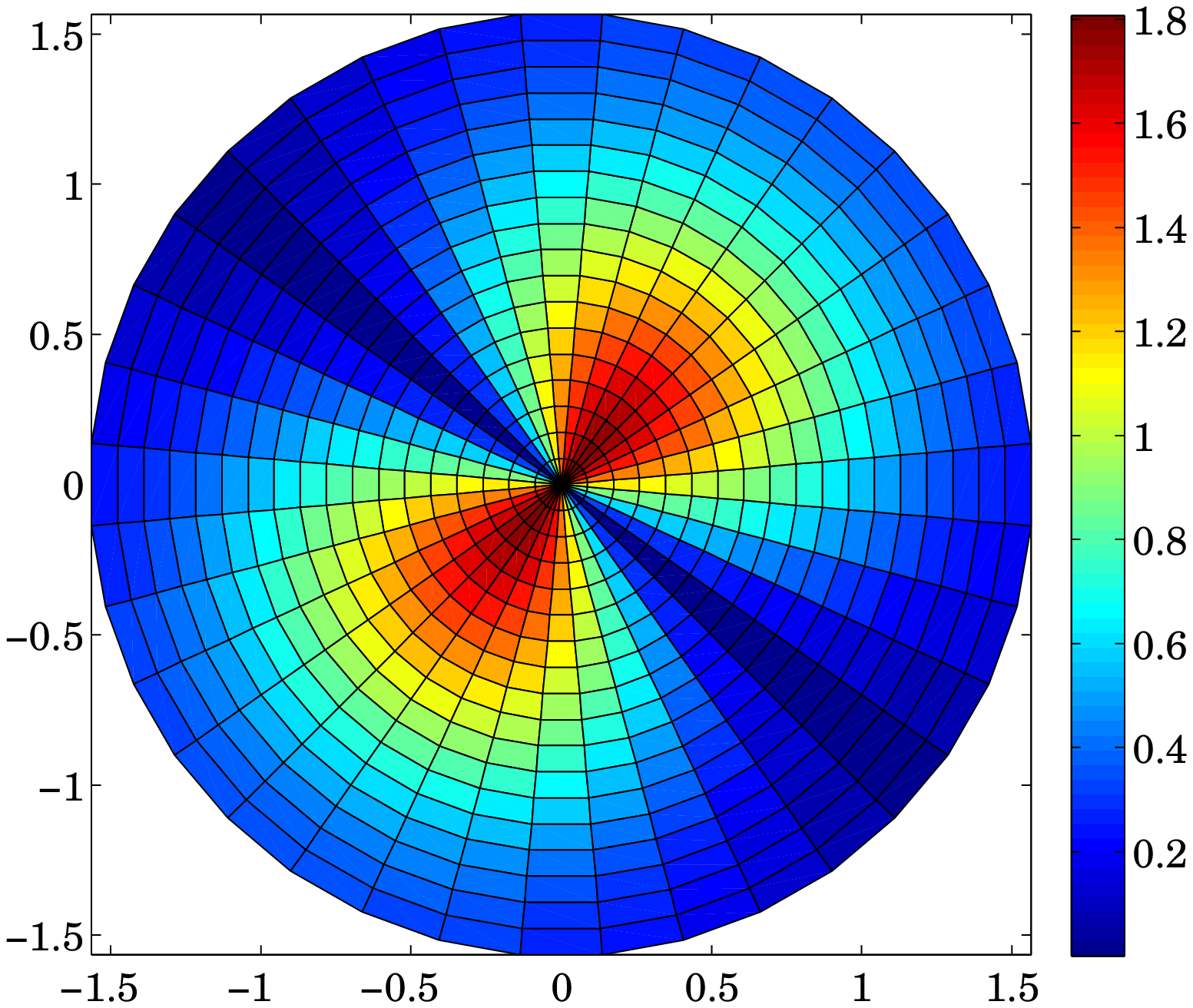,width=4.2cm}}
\vspace{0.1cm} \centerline{(a)}\smallskip
\end{minipage}
\begin{minipage}{0.50\linewidth}
\centering
 \centerline{\epsfig{figure=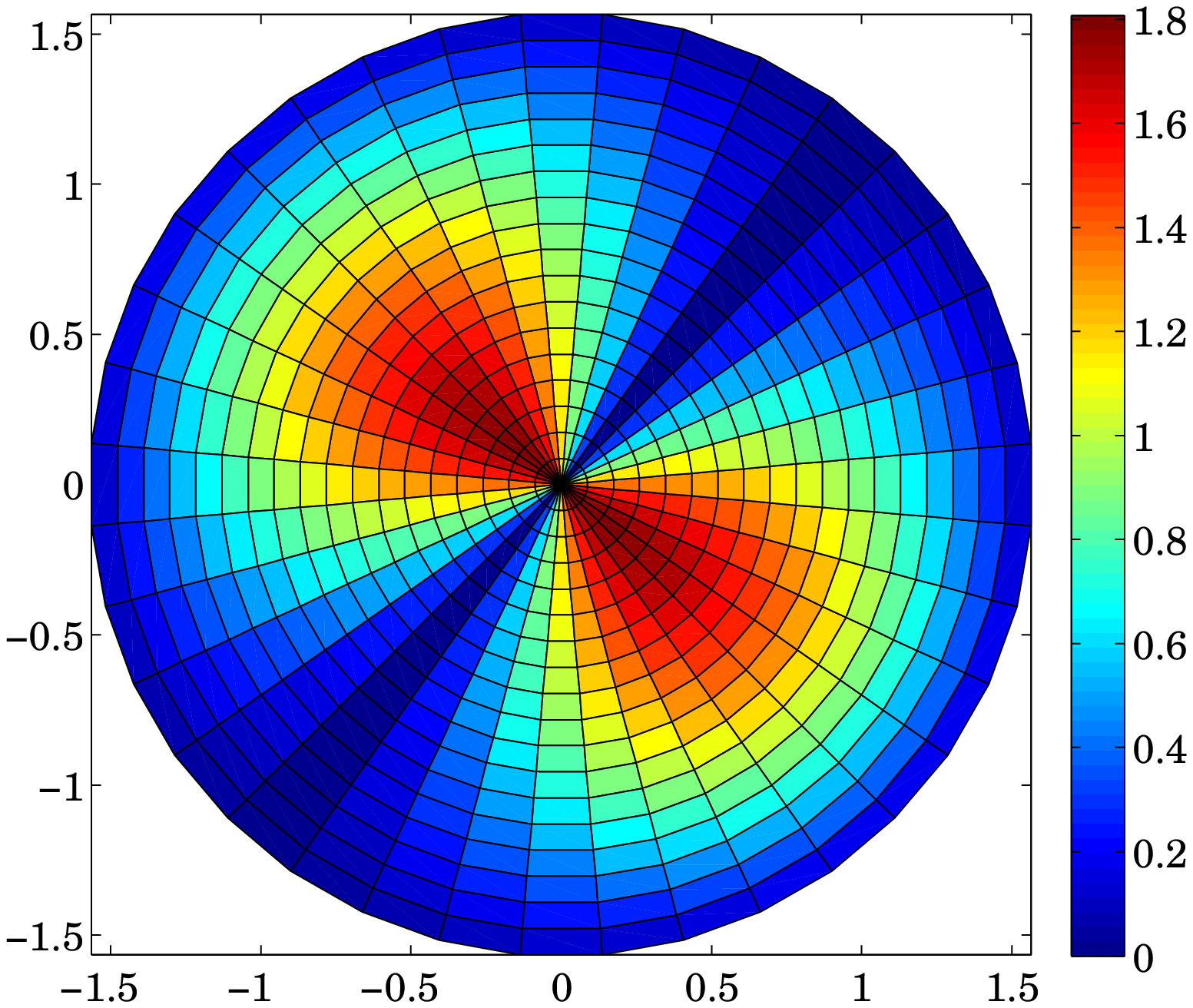,width=4.2cm}}
\vspace{0.1cm} \centerline{(b)}\smallskip
\end{minipage}
\end{minipage}
\caption{The $X$ beam amplitude at $120$ MHz. (a) amplitude $|E_{x\theta}|$ (b) amplitude $|E_{x\phi}|$\label{Ex}}
\end{figure}

\begin{figure}[htbp]
\begin{minipage}{0.99\linewidth}
\begin{minipage}{0.50\linewidth}
\centering
 \centerline{\epsfig{figure=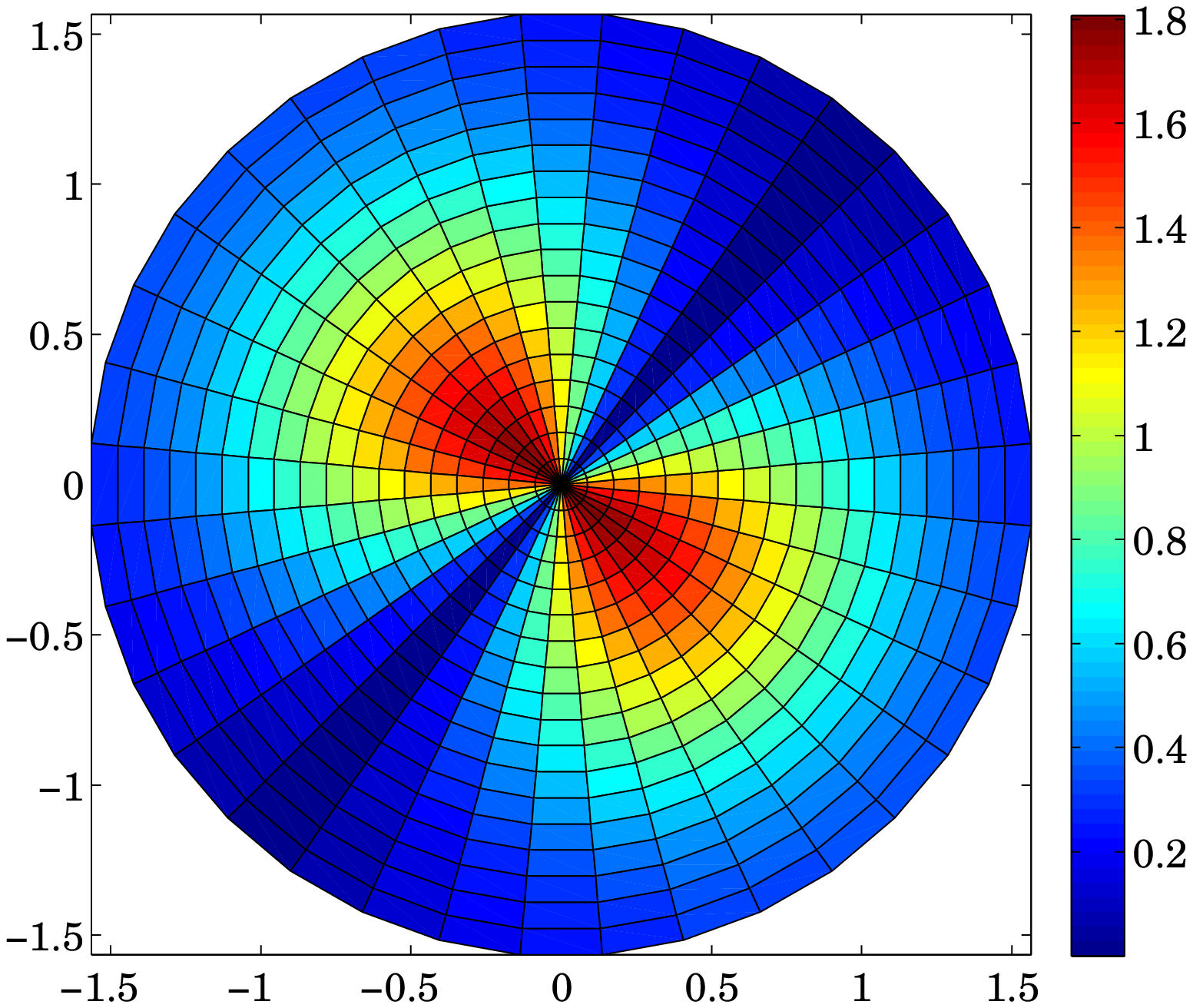,width=4.2cm}}
\vspace{0.1cm} \centerline{(a)}\smallskip
\end{minipage}
\begin{minipage}{0.50\linewidth}
\centering
 \centerline{\epsfig{figure=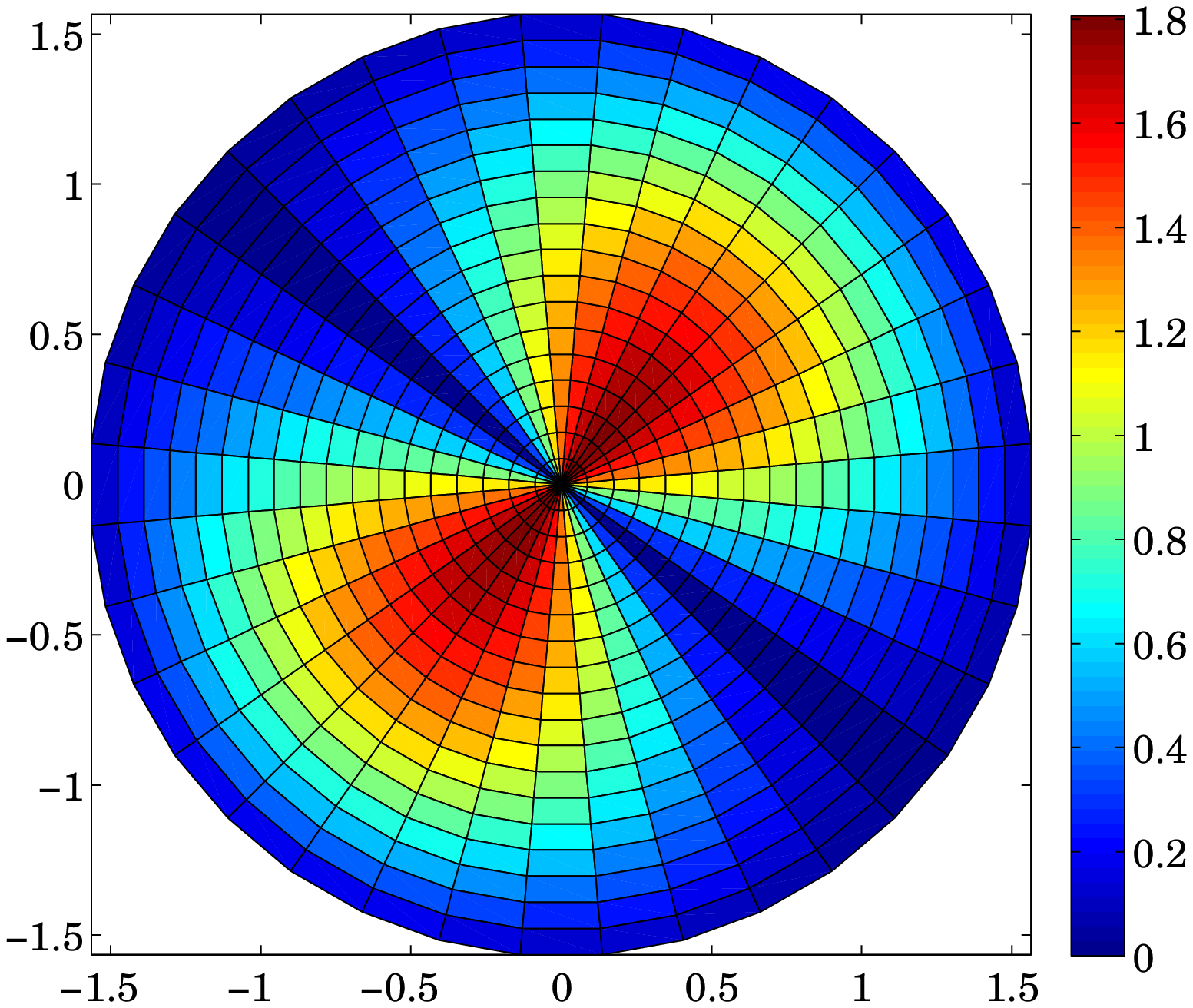,width=4.2cm}}
\vspace{0.1cm} \centerline{(b)}\smallskip
\end{minipage}
\end{minipage}
\caption{The $Y$ beam amplitude at $120$ MHz. (a) amplitude $|E_{y\theta}|$ (b)  amplitude $|E_{y\phi}|$\label{Ey}}
\end{figure}

\begin{figure}[htbp]
\begin{minipage}{0.99\linewidth}
\begin{minipage}{0.50\linewidth}
\centering
 \centerline{\epsfig{figure=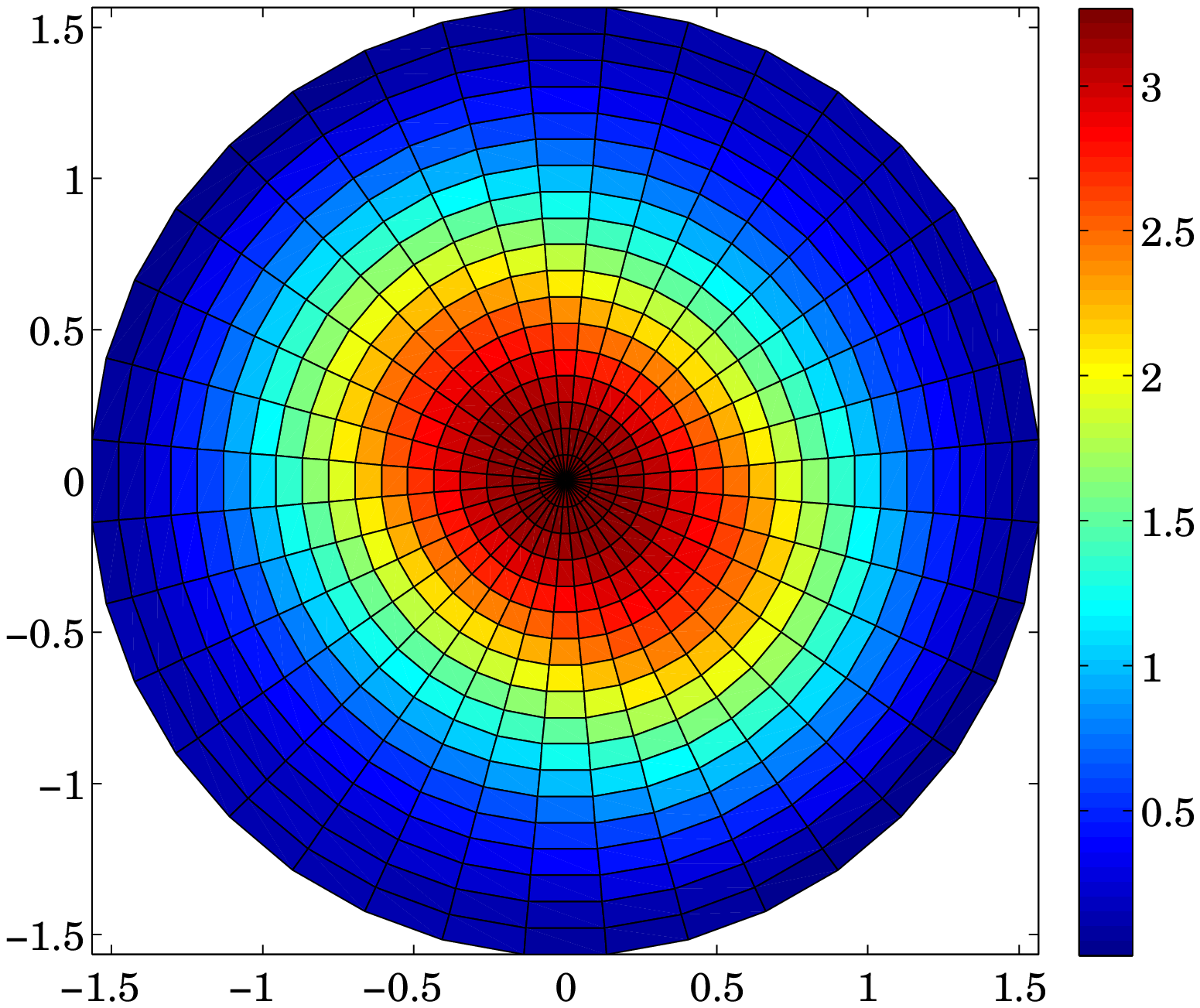,width=4.2cm}}
\vspace{0.1cm} \centerline{(a)}\smallskip
\end{minipage}
\begin{minipage}{0.50\linewidth}
\centering
 \centerline{\epsfig{figure=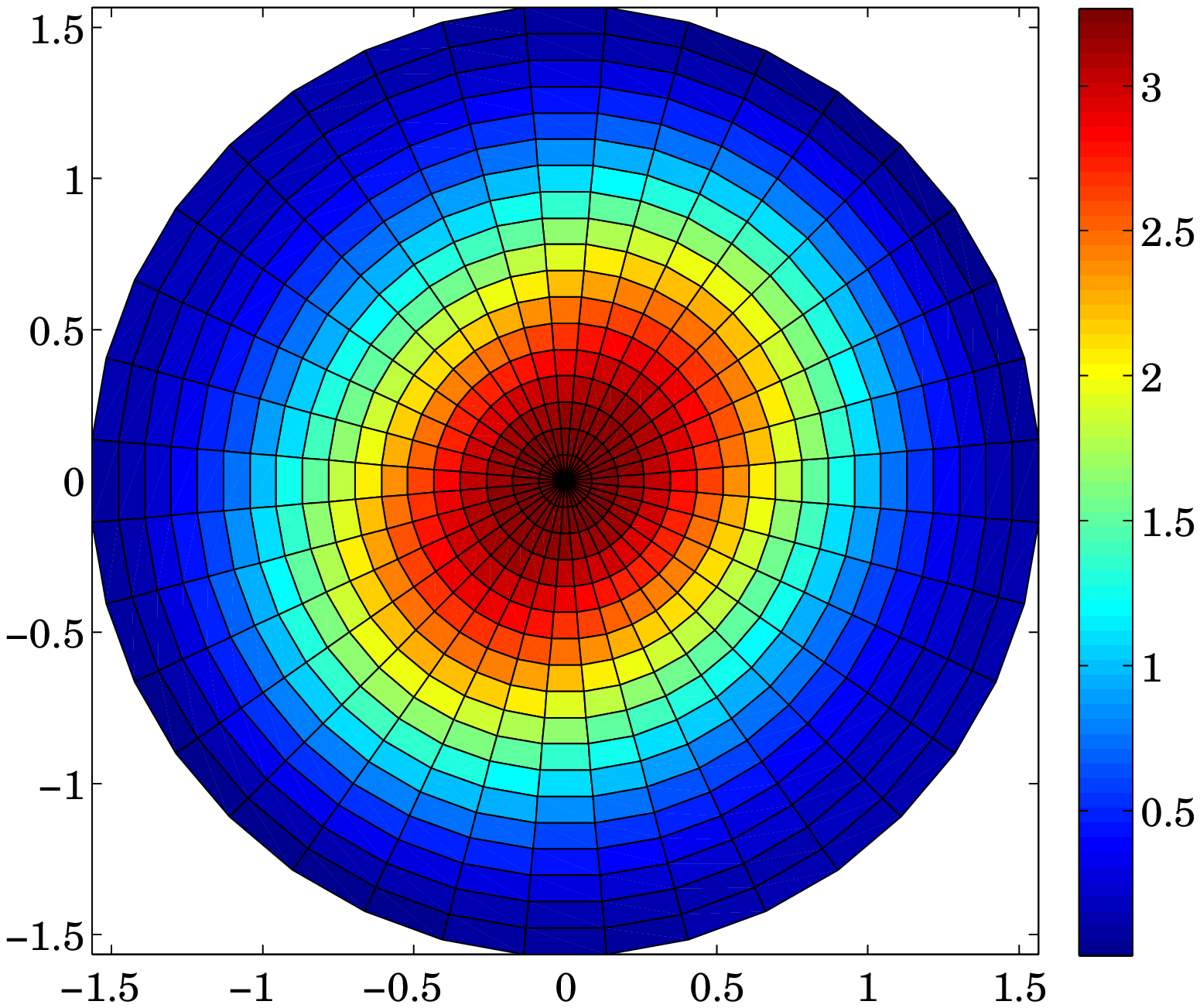,width=4.2cm}}
\vspace{0.1cm} \centerline{(b)}\smallskip
\end{minipage}
\end{minipage}
\caption{The power beam for $X$ and $Y$ dipoles at $120$ MHz. (a) $X$ power beam $|E_{x\theta}|^2 + |E_{x\phi}|^2 $ (b) $Y$ power beam $|E_{y\theta}|^2+|E_{y\phi}|^2$\label{Ep}}
\end{figure}

\begin{figure}[htbp]
\begin{minipage}{0.99\linewidth}
\begin{minipage}{0.50\linewidth}
\centering
 \centerline{\epsfig{figure=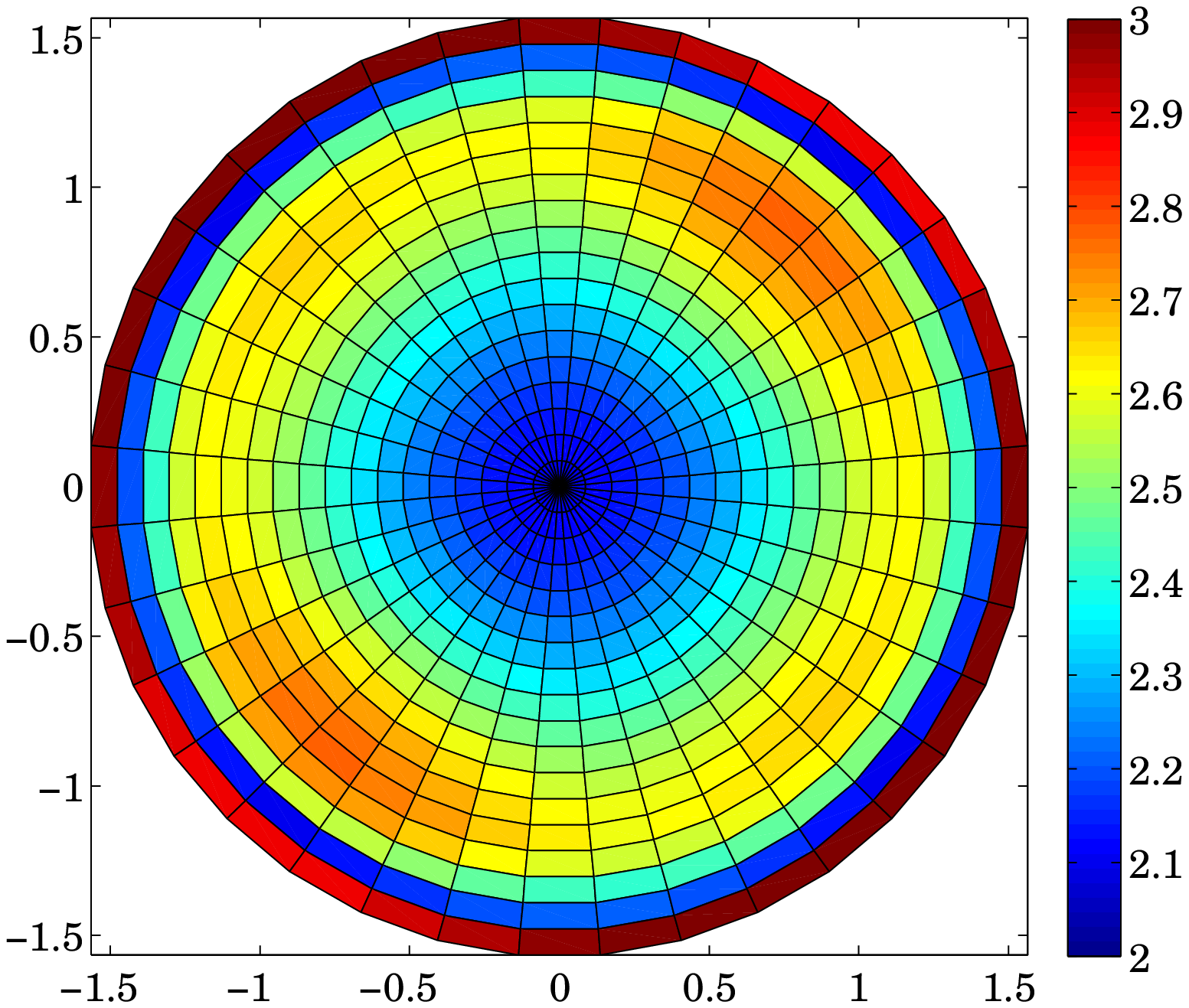,width=4.2cm}}
\vspace{0.1cm} \centerline{(a)}\smallskip
\end{minipage}
\begin{minipage}{0.50\linewidth}
\centering
 \centerline{\epsfig{figure=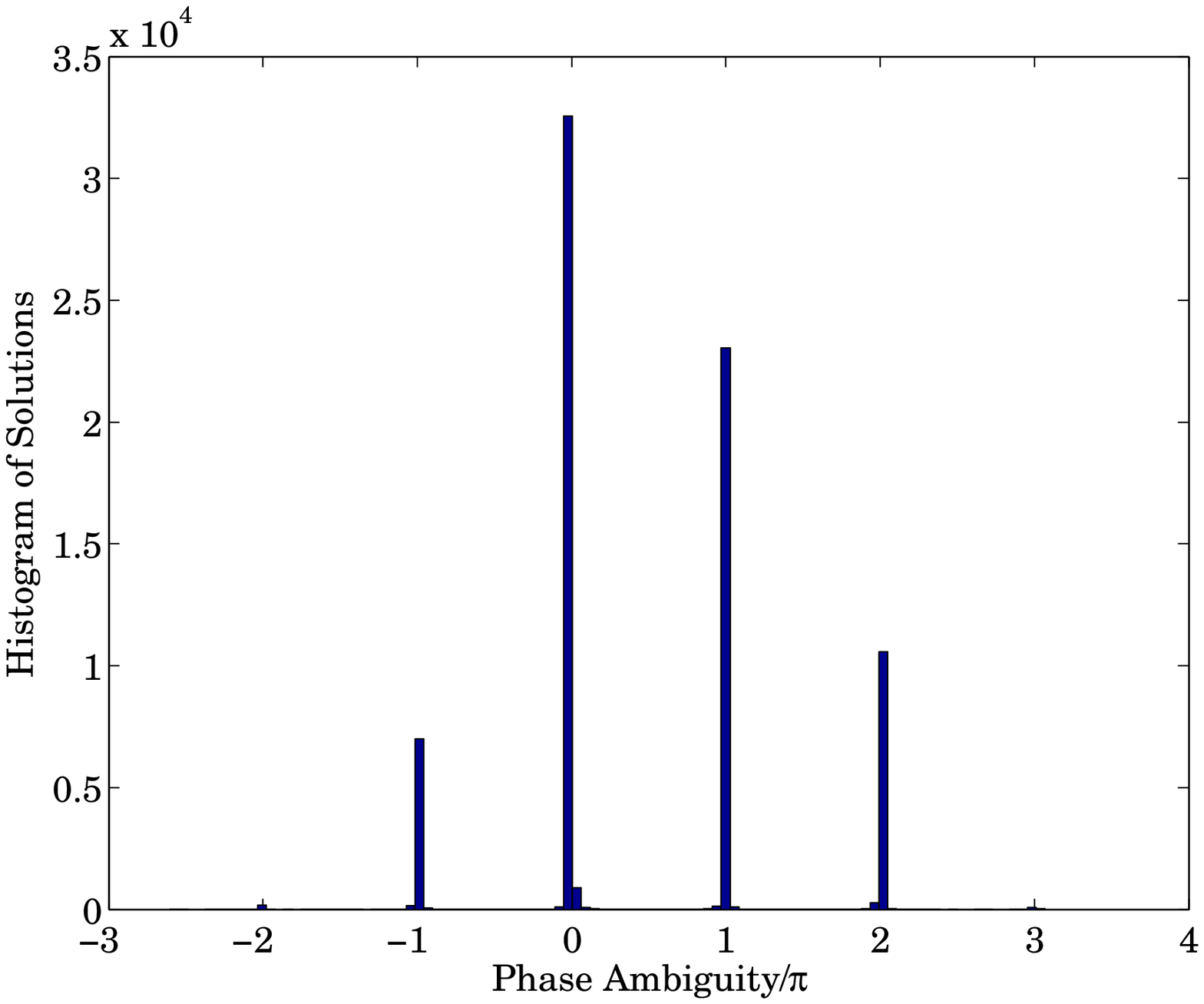,width=4.2cm}}
\vspace{0.1cm} \centerline{(b)}\smallskip
\end{minipage}
\end{minipage}
\caption{(a) Relative condition number (\ref{cond}). (b) Histogram of the error of the solutions for $\phi_{xp}$ and $\phi_{yp}$, divided by $\pi$.\label{frech_sol}}
\end{figure}

We present some results based on numerical simulations in this section.
 We select the observation frequency to be at $120$ MHz. The ${\bf E}$ Jones matrix is derived from numerical electromagnetic simulation of a LOFAR high band antenna (HBA) dipole. The elements of the ${\bf E}$ Jones matrix at this frequency is shown in Fig. \ref{Ex} and Fig. \ref{Ey}.  Both these figures show the beam over the full hemisphere, orthographically projected onto the plane. The dipoles are rotated by $\pi/4$ radians from the meridian. The center of each figure correspond to the zenith. The perimeter of the figure correspond to the horizon. Note that in all figures in this section we use the same projection and coordinate system (except in Fig. \ref{frech_sol} (b)).
In Fig. \ref{Ep}, we have also shown the power beam for both $X$ and $Y$ dipoles.


\begin{figure}[htbp]
\begin{minipage}{0.99\linewidth}
\begin{minipage}{0.48\linewidth}
\centering
 \centerline{\epsfig{figure=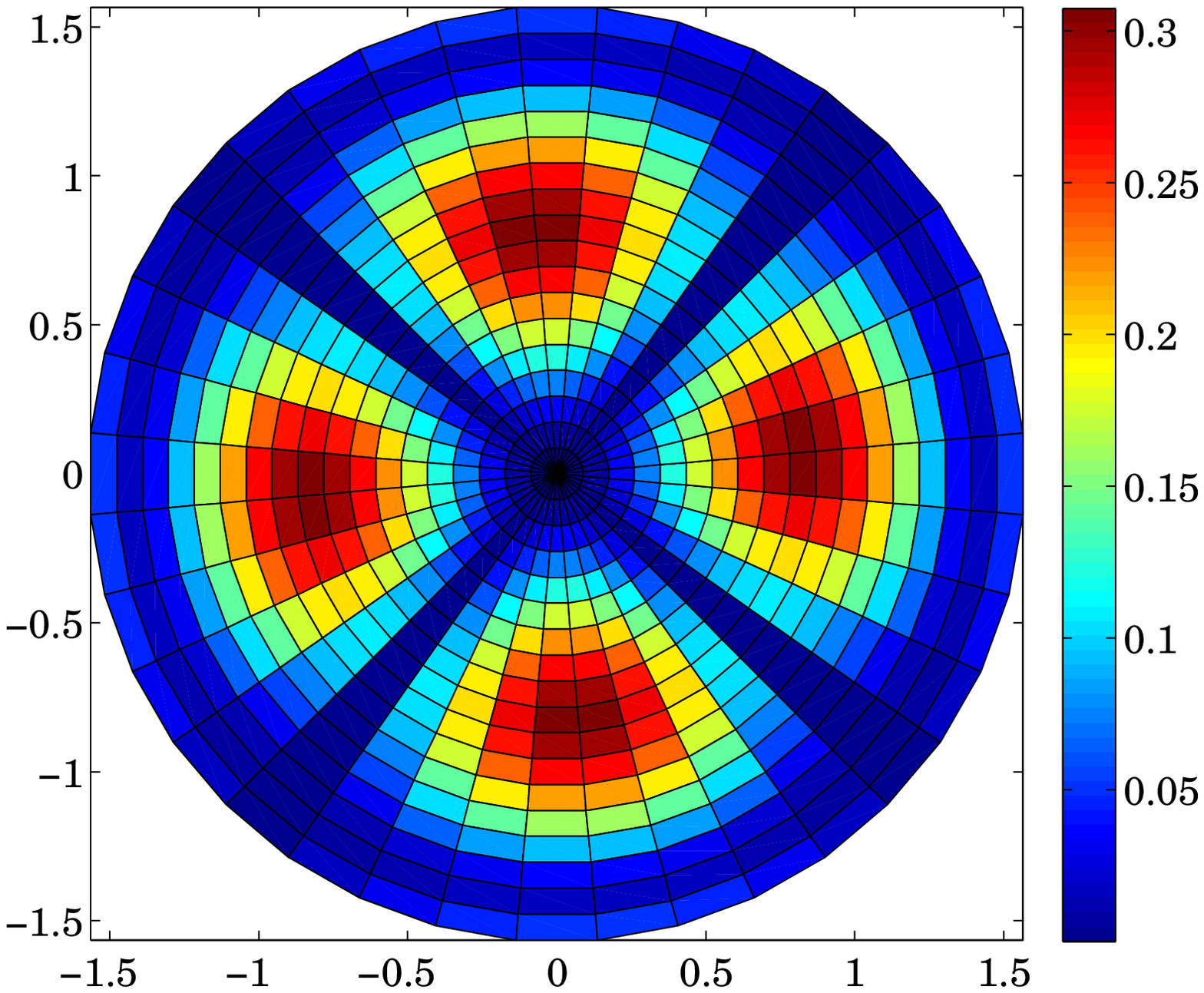,width=4.2cm}}
\vspace{0.1cm} \centerline{(a)}\smallskip
\end{minipage}
\begin{minipage}{0.48\linewidth}
\centering
 \centerline{\epsfig{figure=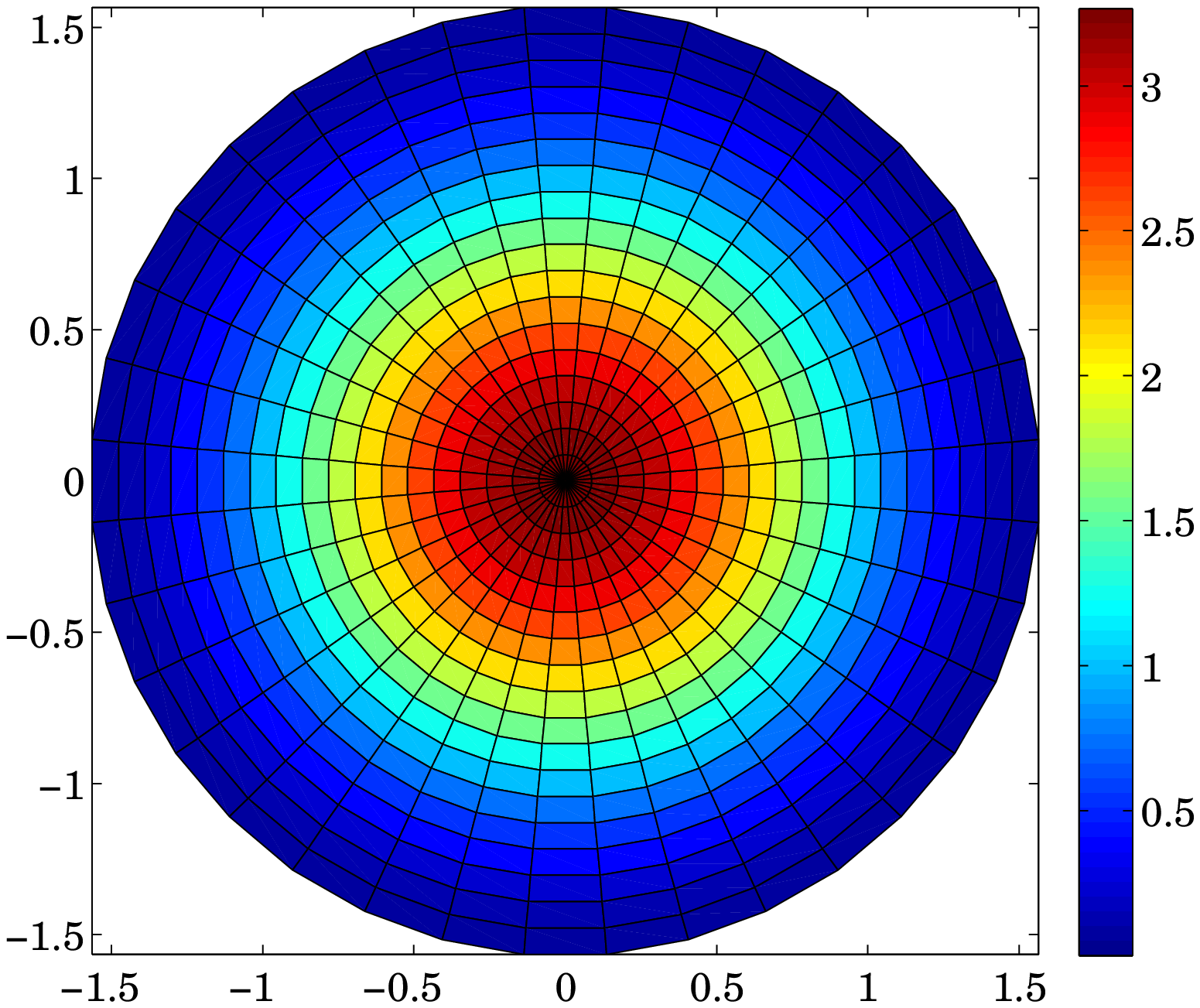,width=4.2cm}}
\vspace{0.1cm} \centerline{(b)}\smallskip
\end{minipage}
\begin{minipage}{0.93\linewidth}
\centering
 \centerline{\epsfig{figure=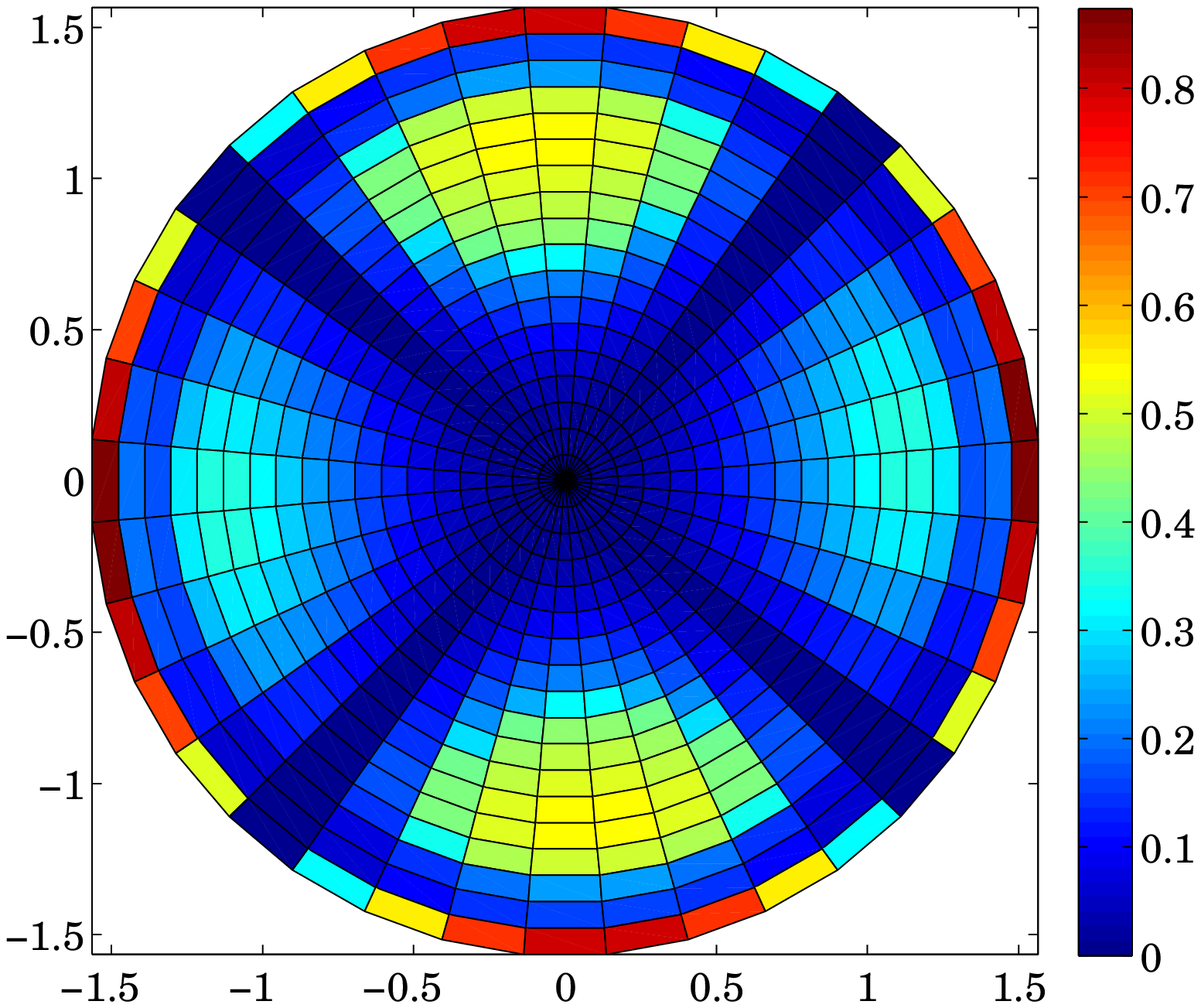,width=4.2cm}}
\vspace{0.1cm} \centerline{(c)}\smallskip
\end{minipage}
\end{minipage}
\caption{Conditions for (\ref{EE}) to be not diagonal. (a) amplitude $|(E_{x\theta}E^{\star}_{y\theta}+ E_{x\phi}E^{\star}_{y\phi})|$ (b) amplitude $|(E_{x\theta}E^{\star}_{y\phi}-E_{x\phi}E^{\star}_{y\theta})|$ (c) amplitude of smallest eigenvalue of $\widetilde{\bf E}$\label{Exp}} 
\end{figure}

Now, let us reconsider the necessary conditions discussed in section \ref{Opt}.  In Fig. \ref{Exp}, we have shown the conditions when (\ref{diagE}) could be satisfied. We have shown the products $|(E_{x\theta}E^{\star}_{y\theta}+ E_{x\phi}E^{\star}_{y\phi})|$ and  $|(E_{x\theta}E^{\star}_{y\phi}-E_{x\phi}E^{\star}_{y\theta})|$ in Fig. \ref{Exp} (a) and (b), respectively. The first term goes close to zero along the planes of the dipoles and at the horizon. However, the second term only goes to zero at the horizon.

In Fig. \ref{Exp} (c), we have shown magnitude of the smallest eigenvalue of $\widetilde{\bf E}$ (normalized by $\|\widetilde{\bf E}\|$) given in  (\ref{matdiagE1}). This figure tells us something more than the previous two figures: which is that apart from the directions along the planes of the dipoles (which also includes the zenith), no possible Faraday rotation exists such that (\ref{diagE}) is satisfied. In other words, regardless of the Faraday rotation, we should be able to estimate it in all directions except the planes of the dipole and the horizon, because (\ref{EE}) is not diagonal.

So far, we have studied the characteristics of the beam that enables our calibration to work. Now, we shall consider the full optimization problem and evaluate its performance. We have simulated an array with $N=54$ LOFAR stations. 
We vary the direction of the celestial source in the sky (keeping the flux at 1 Jy), and for each direction, we do a simulation. The direction is varied at discrete intervals with azimuth range $[0,2\pi)$ radians and elevation range $[0,\pi/2]$ radians.
The gains $|g_{xp}|$,$|g_{yp}|$ are simulated from a uniform distribution in $[0.5,1.5]$. The phases $\phi_{xp}$,$\phi_{yp}$  are simulated from a uniform distribution in $[0,2\pi)$ radians. The Faraday rotations $\alpha_p$ are also simulated from a uniform distribution in $[0,2\pi)$ radians. We have also added white Gaussian noise to the simulated visibilities, keeping the signal to noise ratio at 10. We use a stand alone optimization routine \citep{levmar} to solve the calibration problem (\ref{theta}). The initial values for $|g_{xp}|$,$|g_{yp}|$ are kept at $1$ and the rest of parameters are kept at $0$.
\begin{figure}[htbp]
\begin{minipage}{0.99\linewidth}
\begin{minipage}{0.50\linewidth}
\centering
 \centerline{\epsfig{figure=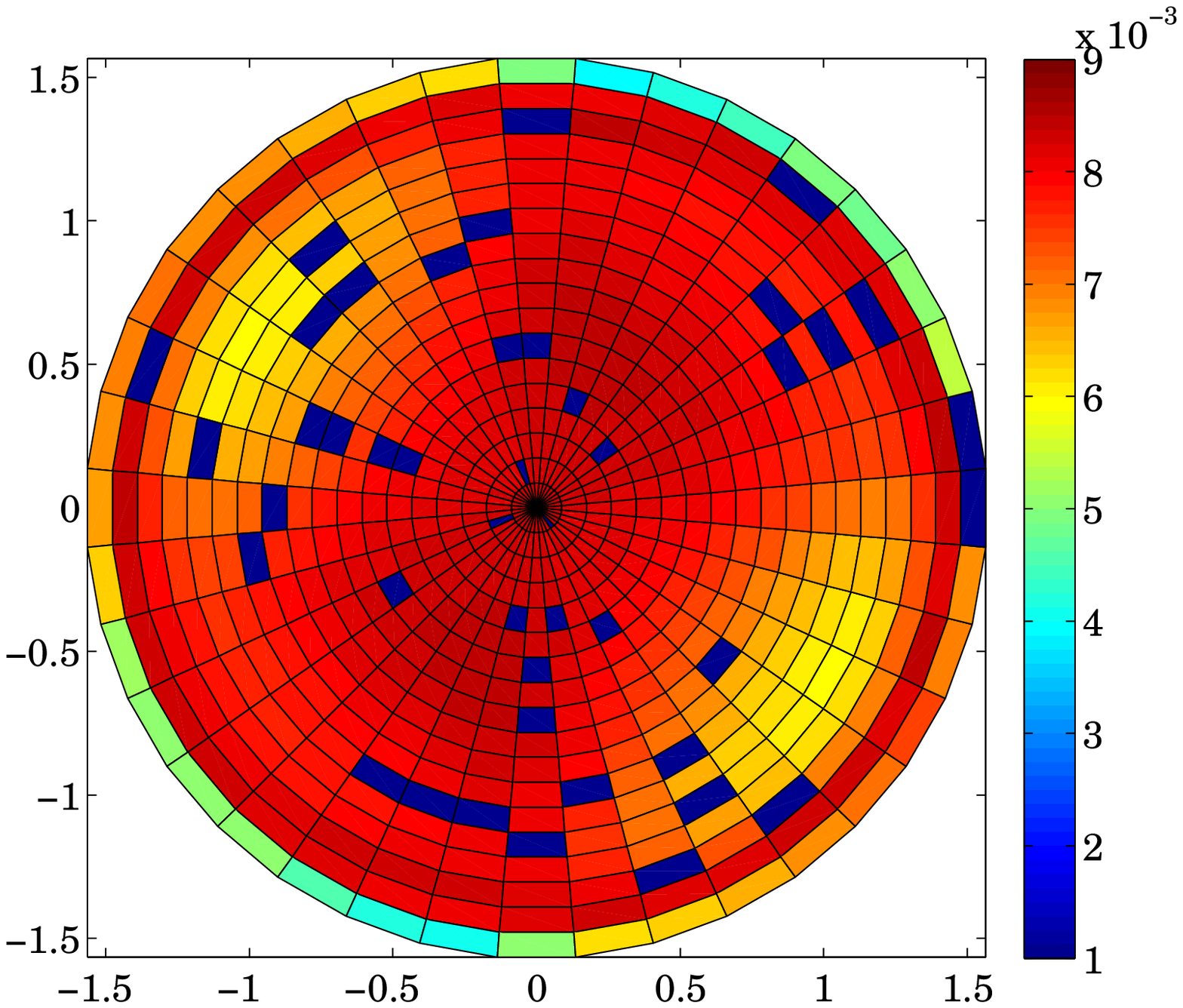,width=4.2cm}}
\vspace{0.1cm} \centerline{(a)}\smallskip
\end{minipage}
\begin{minipage}{0.50\linewidth}
\centering
 \centerline{\epsfig{figure=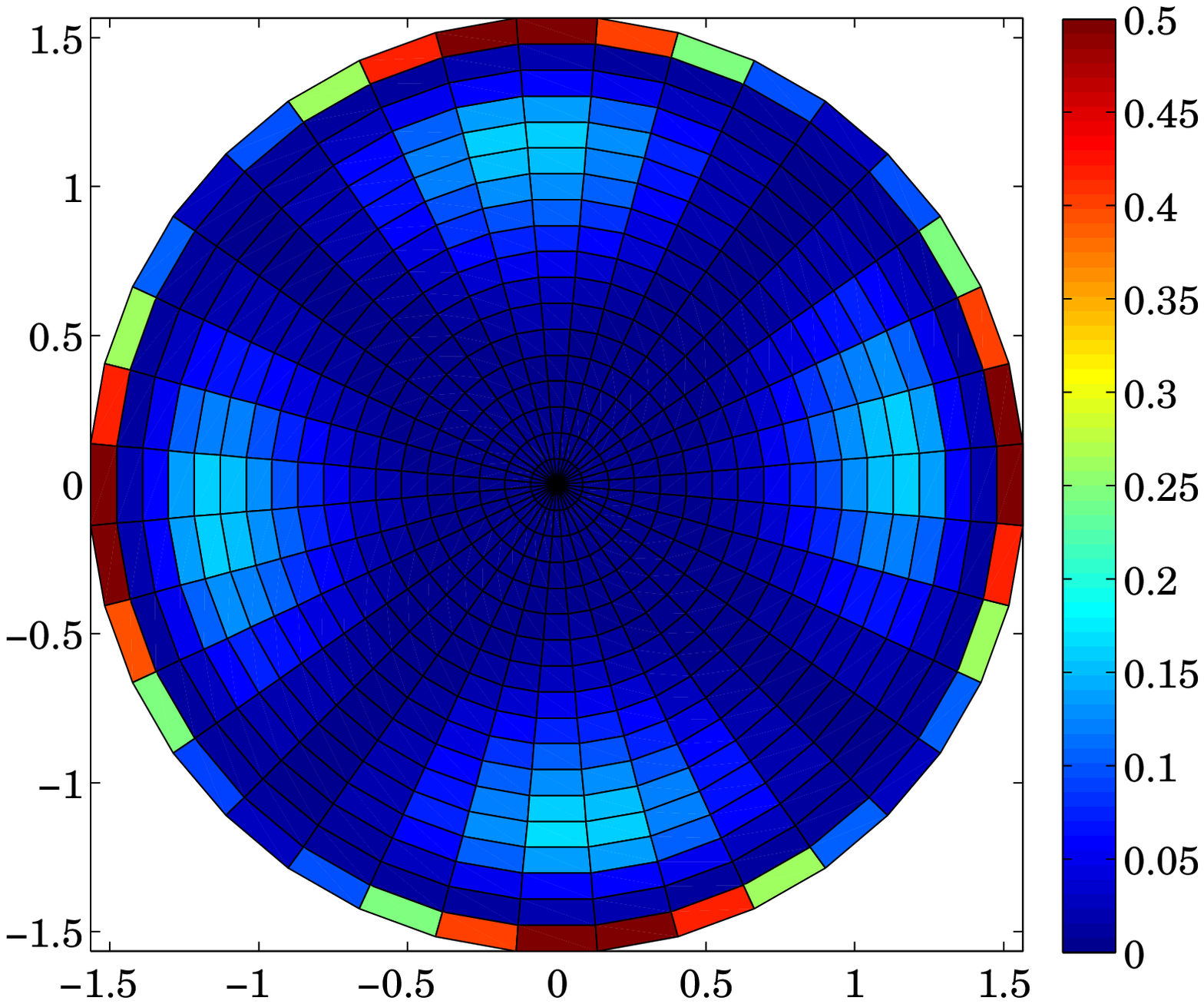,width=4.2cm}}
\vspace{0.1cm} \centerline{(b)}\smallskip
\end{minipage}
\end{minipage}
\caption{(a) The second smallest eigenvalue of the Hessian of (\ref{theta}). (b) Residual normalized error of (\ref{theta}).\label{second}}
\end{figure}

In Fig. \ref{second} (a), we have shown the second smallest eigenvalue of the Hessian obtained after calibration. This is positive for most directions in the sky and where it goes close to zero, we need to exploit diversity to obtain a complete solution. However, the overall behavior shows that the problem is mostly convex. In Fig. \ref{second} (b), we have shown the normalized residual error of (\ref{theta}), which is small and mostly dominated by noise, except for some high values at the horizon.

In Fig \ref{frech_sol} (a), we have shown the relative condition number obtained using (\ref{cond}). The average value of the condition number over all baselines is shown. It is below $3$ for this particular case, indicating that the problem is well conditioned and quite robust to error in modeling of ${\bf E}$. Next, in Fig. \ref{frech_sol} (b), we have shown the histogram of the errors in estimating $\phi_{xp}$ and $\phi_{yp}$. The true error is small and hardly visible in this figure. The dominant feature is the ambiguity in $\pi$. We see similar behavior in the solutions for $\alpha_p$.

\section{Conclusions}
We have shown that it is indeed possible to calibrate LOFAR with reduced ambiguities. The analysis was done for an unpolarized point source, and it is straight forward to extend this to an unpolarized extended source. The crucial factor in our analysis was that the a priori knowledge of the element beam shape. Extensive electromagnetic simulations has in fact given us that knowledge. We have also shown under which conditions our calibration fails to give an answer. However, in such situations, we can exploit diversity to get a complete solution. The extension of this method to polarized sources as well as results based on application of the proposed method to real LOFAR data will be presented in future work.

\section{Acknowledgments}
We thank Ger de Bruyn and Wim Brouw for initial comments and suggestions. We also thank the anonymous reviewers for the careful review and helpful comments that enabled us to enhance this paper.
\bibliographystyle{spbasic}

\begin{thebibliography}{10}
\providecommand{\natexlab}[1]{#1}
\providecommand{\url}[1]{{#1}}
\providecommand{\urlprefix}{URL }
\expandafter\ifx\csname urlstyle\endcsname\relax
  \providecommand{\doi}[1]{DOI~\discretionary{}{}{}#1}\else
  \providecommand{\doi}{DOI~\discretionary{}{}{}\begingroup
  \urlstyle{rm}\Url}\fi
\providecommand{\eprint}[2][]{\url{#2}}

\bibitem[{Boyd and Vendenberghe(2004)}]{CVX}
Boyd S, Vendenberghe L (2004) {Convex optimization}. Cambridge U.K.:Cambridge
  University Press

\bibitem[{Chen and Petropulu(2001)}]{Chen01}
Chen B, Petropulu AP (2001) {Frequency domain blind MIMO system identification
  based on second- and higher order statistics}. IEEE Trans on Signal
  Processing 49(1):1677--1688

\bibitem[{Cotton(1995)}]{Cotton95}
Cotton WD (1995) {Polarimetry}. ASP Conference Series, Very Long Baseline
  Interferometry and the VLBA, JA Zensus, PJ Diamond and PJ Napier (eds)
  82:289--308

\bibitem[{Hamaker(2000)}]{H4}
Hamaker JP (2000) {Understanding Radio Polarimetry IV: The full-coherency
  analogue of scalar selfcalibration}. Astronomy and Astrophysics Supp
  143(3):515--534

\bibitem[{Hamaker et~al(1996)Hamaker, Bregman, and Sault}]{HBS}
Hamaker JP, Bregman JD, Sault RJ (1996) {Understanding radio polarimetry, paper
  I}. Astronomy and Astrophysics Supp 117(137):96--109

\bibitem[{Higham(2008)}]{higham}
Higham NJ (2008) {Functions of matrices: Theory and computation}. Philadelphia
  USA: SIAM

\bibitem[{Intema et~al(2009)Intema, van~der Tol, Cotton, Cohen, van Bemmel, and
  Rottgering}]{Int09}
Intema HT, van~der Tol S, Cotton WD, Cohen AS, van Bemmel IM, Rottgering HJA
  (2009) {onospheric calibration of low frequency radio interferometric
  observations using the peeling scheme. I. Method description and first
  results}. A\&A 501(3):1185--1205

\bibitem[{Lourakis(2004)}]{levmar}
Lourakis MIA (2004) {levmar: Levenberg-Marquardt nonlinear least squares
  algorithms in {C}/{C}++}. \url{http://www.ics.forth.gr/~lourakis/levmar/}

\bibitem[{van~der Tol and van~der Veen(2007)}]{Tol07}
van~der Tol S, van~der Veen AJ (2007) {Ionospheric Calibration for the LOFAR
  Radio Telescope}. IEEE International Symposium on Signals, Circuits and
  Systems, 2007 ISSCS 2007 pp 1--4

\bibitem[{Yatawatta et~al(2004)Yatawatta, Petropulu, and Dattani}]{vehi03}
Yatawatta S, Petropulu AP, Dattani R (2004) {Blind channel estimation using
  fractional sampling}. IEEE Trans on Vehicular Technology 53(2):363--371

\end{thebibliography}

\end{document}